\documentclass[aps,pra,preprintnumbers,amsmath,amssymb,superscriptaddress,floatfix,twocolumn]{revtex4-1}

\usepackage{graphicx}
\usepackage{indentfirst}
\usepackage{subfigure}
\usepackage{color,soul}
\usepackage{braket}
\usepackage{slashbox}
\usepackage[toc,page]{appendix}
\begin{document}

\title{Key rate enhancement using qutrit states for uncharacterized quantum key distribution}
\author{Yonggi Jo}\affiliation{Department of Physics, Sogang University, 35, Baekbeom-ro, Mapo-gu, Seoul 04107, Republic of Korea}
\author{Wonmin Son}\affiliation{Department of Physics, Sogang University, 35, Baekbeom-ro, Mapo-gu, Seoul 04107, Republic of Korea}
\affiliation{Department of Physics, University of Oxford, Parks Road, Oxford OX13PU, United Kingdom}

\begin{abstract}
It is known that measurement-device-independent quantum key distribution (MDI-QKD) provides ultimate security from all types of side-channel attack against detectors at the expense of low key generation rate. Here, we propose MDI-QKD using 3-dimensional quantum states and show that the protocol improves the secret key rate under the analysis of mismatched-basis statistics. Specifically, we analyze security of the 3$d$-MDI-QKD protocol with uncharacterized sources, meaning that the original sources contain unwanted states instead of expected one. We simulate secret key rate of the protocol and identify the regime where the key rate is higher than the protocol with the qubit MDI-QKD.
\end{abstract}

\maketitle

\section{Introduction}

Quantum cryptography is a matured application of quantum informational theory which exploits quantum mechanical principle.  In its core, there is a process called quantum key distribution (QKD) \cite{Bennett84,Ekert91} generating secure key between two distant parties, Alice and Bob, against a possible eavesdropper called Eve. There were many researches for proving security of QKD based on quantum mechanical effects \cite{Deutsch96,Mayer01,Shor00,Devetak05,Renner05} and experiments for demonstrating QKD system \cite{Muller94,Muller95,Muller97,Naik00,Tittel00,Ursin07}.

It is notable that the early QKD protocols use two dimensional quantum state, called qubit \cite{Schumacher95}. Due to its extensible structure, it is expected that high dimensional quantum state is able to carry more information per single quanta compared with qubit. Till now, high dimensional quantum systems on photon were studied for quantum communication in a various context. There were theoretical ideas to exploit high dimensional quantum states in quantum information processing, for example, nonlocality test \cite{Collins02,Son04}, entanglement measurement \cite{Rungta00} and quantum teleportation \cite{Braunstein00,Son01}. Experimentally, high dimensional quantum states are demonstrated in various quantum systems, energy-time entangled states \cite{Thew04,Khan06}, position and momentum entangled states \cite{Neves05,Hale05}, multi path-entangled states \cite{Schaeff12}, and orbital angular momentum(OAM) mode \cite{Leach02,Malik16}.
High dimensional quantum states are applied in QKD protocol as well. There were researches for proving security of QKD using $d$-dimensional quantum system which is generalized version of original QKD protocol \cite{Cerf02,Bourennane02,Durt04,Sheridan10,Ferenczi12,Coles16}. These results show that QKD using high dimensional quantum states has higher upper bound on the error rate that ensures unconditional security of the channel. QKD protocols using high dimensional quantum states are demonstrated by using time-energy states \cite{Pasquinucci00,Khan07,Mower13,Nunn13}, spatial modes \cite{Walborn06,Etcheverry13}, and OAM modes \cite{Groblacher06,Mirhosseini15}, so far.

In the other side of QKD investigation, security of practical QKD system has also been scrutinized in detail. Many theoretical security proofs have been made under the assumptions that all devices are trusted or well characterized for perfect security. However, in a real situation, it becomes necessary to inspect the case that untrusted devices are used seriously because they may be produced by eavesdropper or they just cannot be operated as expected. On the other hand, attack models which exploit imperfect devices, called side channel attack, are proposed and demonstrated recently. They were photon number splitting (PNS) attack \cite{Brassard00}, faked-state attack \cite{Makarov05}, detector efficiency mismatch attack \cite{Makarov06}, detector blinding attack \cite{Lydersen10}, time-shift attack \cite{Qi07}, and laser damage attack \cite{Bugge14}. From the study of hacking QKD system, it is known that imperfection of devices in the system brings serious security problems. There are security patches for each of the attacks, for example using decoy states for preventing PNS attack \cite{Hwang03}, but we need to defend all attacks even including undiscovered one for ensuring perfect security of QKD.

In order to extend the notion of ultimate security, device-independent QKD (DI-QKD) protocol is proposed in 2007 \cite{Acin06,Acin07}. In the DI-QKD study, they proposed a secure QKD scheme that is independent of the device imperfection while the security is guaranteed by nonlocal correlation identified by Clauser-Horne-Shimony-Holt (CHSH) type inequality \cite{Clauser69}. However, it has been turned out that DI-QKD is not easy to be implemented in practice because it requires high quality entanglement source, low-loss against noisy channel and highly efficient detectors. Compensating the practicality, measurement-device-independent QKD (MDI-QKD) protocol is proposed in 2012 \cite{Lo12}. In MDI-QKD scheme, all types of possible side channel attack exploiting imperfection of detectors are overcome by separating detectors from Alice and Bob. For the separation, potentially untrusted third party, called Charlie, is introduced for their QKD. Alice and Bob send their encoded photon to Charlie, then Charlie performs a special type of composite measurement on his incoming photon pair, called Bell state measurement (BSM). After the BSM, Charlie announces the measurement result to Alice and Bob through the classical channel, then the two party establish the correlation between their photon when their encoding bases are same. From the fact that Charlie only act as a referee for the correlation between Alice and Bob, he cannot access to the encoded message as like eavesdropper and it guarantees the unbounded security between Alice and Bob.

Thus, by using MDI-QKD, we can overcome the most of side channel attacks when the major security issues are attributed to the imperfections of detectors \cite{Lo14}. The other advantage of MDI-QKD is that no entanglement is needed like BB84 protocol. There were several experiments to demonstrates MDI-QKD using different physical systems, \textit{e.g.} with time-bin states \cite{Rubenok13} and polarization states \cite{ZTang14}. Especially, the latter realized the long distance MDI-QKD over 200km \cite{YTang14}.

In the mean while, the practical MDI-QKD still suffers from its low key rate compared with BB84 protocol together with high quality requirement that the communication source should satisfy. MDI-QKD needs BSM setup that has only 50\% success probability using linear optical elements \cite{Lutkenhaus99}. Such the success probability of BSM is mainly responsible for a low key generation rate of MDI-QKD.

In this work, we propose MDI-QKD protocol using 3-dimensional quantum state (3$d$-MDI-QKD) which allows the improvement of the secret key rate compared with the original protocol using qubits. In the security analysis, we focus on asymptotic key rates. We analyze the security of 3$d$-MDI-QKD under the assumption that the states generated from communication sources are not ideally prepared, which is called the uncharacterized sources assumption. We use the mismatched-basis statistics in security analysis for 3$d$-MDI-QKD with uncharacterized sources as it is proposed for qubit MDI-QKD in \cite{Yin14}. Here, we show that there is the improvement of the security in 3$d$-MDI-QKD compared with qubit MDI-QKD in theoretical model, even if the communication sources are uncharacterized. We simulate the secret key rate of 3$d$-MDI-QKD with the change of the realistic experimental factors and identify the regime where 3$d$-MDI-QKD is more secure than qubit MDI-QKD.

This article is organized as following. In section \ref{sec3dmdi}, we present the schematic description of 3$d$-MDI-QKD. In section \ref{sec3dmdiuc}, we analyze the security of 3$d$-MDI-QKD with uncharacterized sources by using the analysis of the mismatched-basis statistics. In section \ref{secsimul}, we simulate the secret key rate of 3$d$-MDI-QKD and compare it with that of qubit MDI-QKD in the theoretical model and in the realistic experiment model. Finally, we conclude in section \ref{secconclusion}. Details of calculation are given in the Appendix.

\section{MDI-QKD using 3-dimensional quantum states}\label{sec3dmdi}
\begin{figure}[!ht]
\begin{subfigure}[Time basis]{\includegraphics[width=0.47\textwidth]{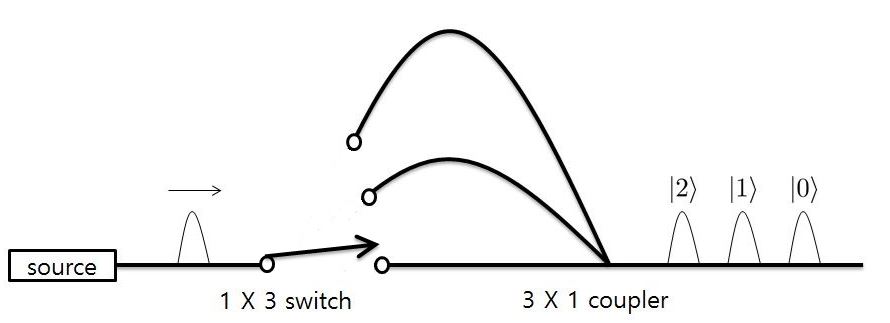}}\end{subfigure}
\begin{subfigure}[Energy basis]{\includegraphics[width=0.47\textwidth]{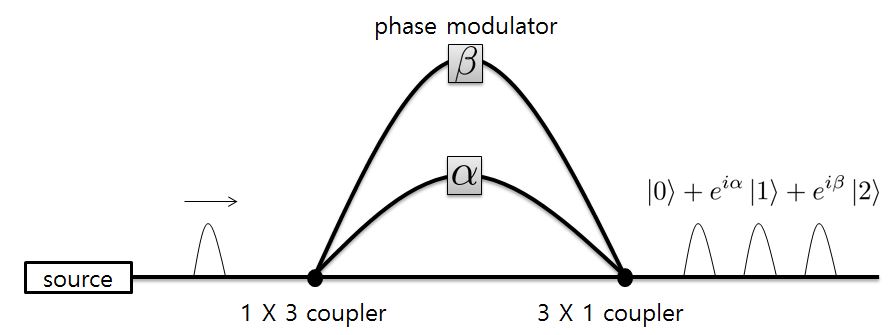}}\end{subfigure}
\caption{A schematic setup to generate 3-dimensional quantum states of two bases\cite{Pasquinucci00}. In time basis (a), Alice and Bob encode information in time states of photon. In this setup, the time states are generated with three different delay lines and switch. The time that photon injected into the output port can be controlled with choice of delay line. $\ket{0}$, $\ket{1}$ and $\ket{2}$ denote time states when photon passes through the shortest, the middle, and the longest delay line respectively. In energy basis (b), Alice and Bob prepare similar settings with phase modulators on delay lines and $1\times 3$ coupler instead of switch. From Eq.~(\ref{mubs}), in order to generate the states in the energy basis, $\{e^{i\alpha},e^{i\beta}\}$ should be $\{1,1\}$, $\{\omega^{2},\omega\}$, and $\{\omega,\omega^{2}\}$ to generate $\ket{\bar{0}}$, $\ket{\bar{1}}$ and $\ket{\bar{2}}$ respectively.}
\label{3dtebase}
\end{figure}

In this section, we present the schematic description of 3$d$-MDI-QKD. As its extension of the original qubit MDI-QKD protocol \cite{Lo12}, we assume that Alice and Bob use two measurements at each site. In the original protocol, they use two orthonormal states to encode classical binary bits 0 and 1. In our protocol, Alice and Bob use two measurements at each site whose bases for the measurement outcomes are consisted of three orthonormal states. We assume that the two different measurements are prepared in mutually unbiased bases (MUBs). The condition for two bases to be MUBs in 3-dimensional Hilbert space is that $|\braket{i|\bar{j}}|^{2}=1/3$ for all $i,j\in\{0,1,2\}$, where $\{\ket{0},\ket{1},\ket{2}\}$ and $\{\ket{\bar{0}},\ket{\bar{1}},\ket{\bar{2}}\}$ are two orthonormal bases.

The 3-dimensional quantum states can be realized through the various photonic degree of freedom in practice \cite{Pasquinucci00, Walborn06, Etcheverry13, Mirhosseini15}. Fig.~\ref{3dtebase} shows schematic setup to generate 3-dimensional quantum states using the different path length of the optical fiber. QKD protocol using the encoding system is originally proposed in \cite{Pasquinucci00}. In the time basis (Fig.~\ref{3dtebase} (a)), Alice controls the time that photon is injected into the output port of the settings. This task can be accomplished by using $1\times 3$ optical switch and the three different delay lines. The arrival time of photon at output port can be controlled since the delay lines have different path length. In Fig.~\ref{3dtebase} (a), $\ket{0}$, $\ket{1}$ and $\ket{2}$ denote the three time states when photon passes through the shortest, the middle, and the longest delay line respectively. The time intervals among these three time states must be large enough compared with pulse duration of source. In that case, these three states can be distinguished by using measurement of arrival time of photon. In order to create states belonging to the energy basis, Alice uses the setup shown in Fig.~\ref{3dtebase} (b). The length of each delay lines should be same with that of the time basis. There are phase modulators and $1\times 3$ coupler instead of $1\times 3$ switch. $1\times 3$ coupler splits a beam of light into three output ports with same probabilities. In that case, the output state of this setup is described by $\frac{1}{\sqrt{3}}\left(\ket{0}+e^{i\alpha}\ket{1}+e^{i\beta}\ket{2}\right)$ where $\alpha$ and $\beta$ are phase factors that Alice can control. Using the operation, three orthogonal states of the other MUB can be generated and they are described as
\begin{align}
\ket{\bar{0}}&=\frac{1}{\sqrt{3}}\left( \ket{0} +\ket{1} +\ket{2} \right)\label{mubs}\\
\ket{\bar{1}}&=\frac{1}{\sqrt{3}}\left( \ket{0} +\omega^{2} \ket{1} +\omega\ket{2} \right) \nonumber\\
\ket{\bar{2}}&=\frac{1}{\sqrt{3}}\left( \ket{0} +\omega \ket{1} +\omega^{2}\ket{2} \right), \nonumber
\end{align}
where $\omega^{3}=1$, $\omega^{2}+\omega+1=0$ \cite{Wootters89}, and $\ket{\bar{0}}$, $\ket{\bar{1}}$ and $\ket{\bar{2}}$ denote the orthonormal states in the energy basis. Throughout this paper, we denote two MUBs for 3$d$-MDI-QKD ordinary basis and bar basis. In the example, the ordinary basis and the bar basis are corresponding to the time basis and the energy basis respectively.

Before introducing 3$d$-MDI-QKD, it is necessary to redefine the representation of Bell state measurement (BSM) as to describe the maximally entangled states of 3-dimensional bipartite system. There are nine maximally entangled states in 3-dimensional bipartite system. We define $\{\ket{\Phi_{i}}\}$ as a set of 3-dimensional maximally entangled states where $i\in\{0,1,...,8\}$, and each state is described as
\begin{align} \label{3dMEstates}
\left | \Phi_{3k+l} \right > &= \frac{1}{\sqrt{3}}\sum_{m=0}^{2}\omega^{ml} \left| m+k,m \right >
\end{align}
where $k,l\in\{0,1,2\}$. We omit (mod 3) from all indices as a matter of simplification. Then the 3-dimensional BSM (3$d$-BSM) is defined as a set of projections $\{\hat{B}_{i}|\hat{B}_{i}=\ket{\Phi_{i}}\bra{\Phi_{i}}\}$.
\begin{figure}[!b]
\includegraphics[width=0.5\textwidth]
{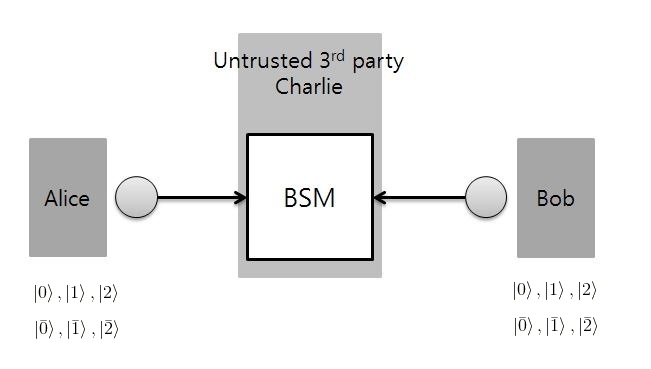}
\caption{A schematic diagram of MDI-QKD using 3-dimensional quantum states (3$d$-MDI-QKD). BSM means Bell-state measurement. In 3-dim MDI-QKD, BSM should be able to discriminate 3-dimensional maximally entangled states.}\label{3dimMDI-QKD}
\end{figure}

\begin{table*}[ht!]
\begin{tabular}{c|c c c c c c c c c}
\hline
\backslashbox{Basis}{BSM result} & $\ket{\Phi_{0}}$ & $\ket{\Phi_{1}}$ & $\ket{\Phi_{2}}$ & $\ket{\Phi_{3}}$ & $\ket{\Phi_{4}}$ & $\ket{\Phi_{5}}$ & $\ket{\Phi_{6}}$ & $\ket{\Phi_{7}}$ & $\ket{\Phi_{8}}$ \\
\hline
Ordinary basis & - & - & - & $\begin{matrix}l\rightarrow\\ l+1\end{matrix}$ &
$\begin{matrix}l\rightarrow\\ l+1\end{matrix}$&
$\begin{matrix}l\rightarrow\\ l+1\end{matrix}$&
$\begin{matrix}l\rightarrow\\ l+2\end{matrix}$ &
$\begin{matrix}l\rightarrow\\ l+2\end{matrix}$ &
$\begin{matrix}l\rightarrow\\ l+2\end{matrix}$ \\
\hline
Bar basis & $1\leftrightarrow2$ &
$0\leftrightarrow2$ &
$0\leftrightarrow1$ &
$1\leftrightarrow2$ &
$0\leftrightarrow2$ &
$0\leftrightarrow1$ &
$1\leftrightarrow2$ &
$0\leftrightarrow2$ &
$0\leftrightarrow1$ \\
\hline
\end{tabular}
\caption{In 3$d$-MDI-QKD protocol, Alice and Bob communicate with each other in order to share the information about the encoding basis. After that, they should do post-process based on result of 3$d$-BSM in order to generate sifted key. In the Table, we represent the post-process method for the case that Alice preserves hers data, and Bob modifies his data. (mod 3) is omitted from the indices and $l\in\{0,1,2\}$.}\label{3dimMDITable1}
\end{table*}

From now on, we discuss the procedure of 3$d$-MDI-QKD. Similarly like the original protocol \cite{Lo12}, Alice (Bob) firstly chooses an integer number, indexed $i$, among 0, 1, 2, \={0}, \={1}, and \={2} randomly. Then she (he) generates the corresponding state $\ket{i}$ and sends it to untrusted third party called Charlie. With the states, Charlie performs 3$d$-BSM on the incoming photons and announces the result of measurement through the public channel. After the measurement, Alice and Bob share the information about the encoding basis through the public channel. Subsequently, after the basis comparison, they discard the trial if  the original bases are different. The remaining data becomes sifted key after post-process based on the result of 3$d$-BSM. In order to synchronize the encoded information, it is necessary to perform the appropriate post-process. The method of post-process is described in Table \ref{3dimMDITable1}.

To evaluate the usefulness of the protocol, it is necessary to analyze security of 3$d$-MDI-QKD. The analysis can be made through the detailed inspection of the equivalent protocol using entanglement distillation process (EDP) \cite{Deutsch96,Shor00,Devetak05}. The idea is that if Alice and Bob share the maximally entangled state, Eve can not generate correlation between her state and the state of Alice and Bob \cite{Coffman00}. The property of entanglement is called monogamy of entanglement. In the sense, QKD is always secure when Alice and Bob share the maximally entangled states. Therefore, in this case, we can analyze security of 3$d$-MDI-QKD with the amount of maximally entangled states between Alice and Bob generated from EDP. The equivalent protocol that exploit the entangled pair can be found as follows.
\begin{enumerate}
 \item Alice and Bob prepare photon pair in a maximally entangled state $\ket{\Phi_{0}}$. Labels of Alice's photons are denoted as A and C, and those of Bob's photons are B and D. In the situation, they start with the pairs of entanglement as
  \begin{align}
  \ket{\Phi_{0}}_{AC}&=\frac{1}{\sqrt{3}}\left(\ket{0,0}_{AC}+\ket{1,1}_{AC}+\ket{2,2}_{AC}\right)\\
  \ket{\Phi_{0}}_{BD}&=\frac{1}{\sqrt{3}}\left(\ket{0,0}_{BD}+\ket{1,1}_{BD}+\ket{2,2}_{BD}\right).\nonumber
  \end{align}
 \item Alice (Bob) sends photon C (D) to Charlie.
 \item Charlie performs 3$d$-BSM onto the incoming photons, and announces his result of measurement to Alice and Bob.
 \item The photon A and B become one of maximally entangled state after 3$d$-BSM \cite{Zukowski93} if there is no loss and there is no Eve.
 \item In order to obatin $\ket{\Phi_{0}}_{AB}$, Bob does unitary operation on his photon based on the result of 3$d$-BSM.
 \item Alice (Bob) chooses measurement basis and measures hers (his) photon A (B).
 \item Alice and Bob compare their measurement bases and accept the encoded number as a key only when they have used the same measurement bases.
 \item If Alice and Bob used bar bases, Bob do post-process to synchronize information. (See Table \ref{3dimMDITable1})
 \item The remaining data is used as sifted key.
\end{enumerate}
In step 5, Alice and Bob share the maximally entangled state $\ket{\Phi_{0}}_{AB}$ if there is no error and no Eve. Then this entanglement version of 3$d$-MDI-QKD becomes same with 3-dimensional case of $d$-dimensional entanglement based protocol proposed in \cite{Durt04}. Security analysis of the $d$-dimensional entanglement based protocol has been studied in the previous works, against individual attack \cite{Durt04} (Eve monitors states separately) and against collective attack \cite{Ferenczi12,Sheridan10} (Eve monitors several states jointly). According to the results, the secret key rate of 3$d$-MDI-QKD per sifted key against collective attack is evaluated as
\begin{equation}\label{3dQKDrate}
r\geq\log_{2}3-2Q-2H(Q)
\end{equation}
where $H(x)$ is Shannon entropy defined as $H(x)=-x\log_{2}x-(1-x)\log_{2}(1-x)$ and $Q$ denotes state error rate in the ordinary basis. If the state that Alice and Bob share at the end of protocol is not $\ket{\Phi_{i}}$ where $i=0,1,2$ then the trial has the state error in the ordinary basis. Then the state error rate in the ordinary basis is described by
\begin{align}\label{stateerror}
Q=\sum_{i\neq j}\bra{i,j}\hat{\rho}\ket{i,j},
\end{align}
where $\hat{\rho}$ is density operator of Alice and Bob, and $i,j\in\{0,1,2\}$. In the ideal case without eavesdropper, the state is remained to be $\hat{\rho}=\ket{\Phi_{0}}\bra{\Phi_{0}}$, so thus the error rate becomes trivial since $Q=0$.

\section{3$d$-MDI-QKD with uncharacterized sources}\label{sec3dmdiuc}

\begin{table}[b]
\centering
\begin{subtable}[ Ideal case]{
\begin{tabular}{c|c c c c c c c c c}
\hline
\backslashbox{BSM}{$x$,$y$} & 0,0~ &0,1~ & 0,2 ~&1,0~ &1,1~ &1,2~ &2,0 ~& 2,1~ &2,2 ~\\
\hline
$p(x,y)$ & 1/3 & 0 & 0 & 0 & 1/3 & 0 & 0 & 0 & 1/3 \\
\hline\hline
\backslashbox{BSM}{$x$,$y$} & 0,\={0} & 0,\={1} & 0,\={2} & 1,\={0} & 1,\={1} & 1,\={2} & 2,\={0} & 2,\={1} & 2,\={2} \\
\hline
$p(x,y)$ & 1/9 & 1/9 & 1/9 & 1/9 & 1/9 & 1/9 & 1/9 & 1/9 & 1/9 \\
\hline
\end{tabular}}
\end{subtable}\\
\begin{subtable}[ Misalignment case]{
\begin{tabular}{c|c c c}
\hline
\backslashbox{BSM}{$x$,$y$} & 0, $\bar{\beta}_{0}$ & 1, $\bar{\beta}_{0}$ & 2, $\bar{\beta}_{0}$\\
\hline
$p(x,y)$ & $(\cos^2\mu\sin^{2}\nu)/3$ & $(\sin^2\mu\sin^{2}\nu)/3$ & $(\cos^{2}\nu)/3$\\
\hline
\end{tabular}}
\end{subtable}
\caption{The success probabilities of 3$d$-BSM $p(x,y)$ when Alice and Bob send $\ket{x}$ and $\ket{y}$ to Charlie. We assume that 3$d$-BSM is able to discriminate only $\ket{\Phi_{0}}$ and there is no loss and no Eve. The success probabilities of 3$d$-BSM are calculated from Tr$\left[\hat{\rho}_{AB}\ket{\Phi_{0}}\bra{\Phi_{0}}\right]$. (a) In ideal case, Alice and Bob send the state in the ordinary basis or the bar basis. Since these two bases are MUBs, the success probabilities of mismatched-basis cases are same with $1/9$. (b) In misalignment case, we assume that Bob's encoder has misalignment while Alice's is perfect. Bob's source generates $\ket{\bar{\beta}_{0}}$ instead of $\ket{\bar{0}}$, where $\ket{\bar{\beta}_{0}}=\cos\mu\sin\nu\ket{0}+\sin\mu\sin\nu\ket{1}+\cos\nu\ket{2}$ for arbitrary angle $\mu$ and $\nu$. The success probabilities of mismatched-basis cases are modified from $1/9$.}\label{Table3dMBS}
\end{table}

MDI-QKD has an advantage that prevents all type of side channel attack against detectors. It also guarantees the security even if detectors are fabricated or controlled by Eve. In order to ensure the perfect security of MDI-QKD, there must be the assumption that sources used in QKD are ideally prepared without error. The states generated from the sources are well characterized in the assumption. For example, sources should generate one of the states among $\{\ket{H},\ket{V},\ket{+},\ket{-}\}$ in the original MDI-QKD to guarantee the perfect security. Here $\ket{H}$ is a single photon state with horizontal polarization, $\ket{V}$ is a photon with vertical polarization, and $\ket{\pm}=\frac{1}{\sqrt{2}}(\ket{H}\pm\ket{V})$. However, in real QKD setup, communication sources can have misalignment in their encoding system and the states generated from the communication sources can be prepared differently from the wanted state.

Through the analysis of mismatched-basis statistics (MBS), it is proven that QKD with uncharacterized sources can also be secure against possible channel eavesdropping \cite{Yin14}. In the original QKD protocol, the data that is obtained from the mismatched-basis are discarded because Alice and Bob cannot extract any information from it. However, in \cite{Yin14}, they argue that the misalignment of communication sources influence to the MBS such that Alice and Bob can still extract secret key partially with the analysis of MBS even if the sources are uncharacterized.

Table \ref{Table3dMBS} shows that the effect of misalignment in the prepared states modifies the MBS in 3$d$-MDI-QKD. At first, we consider the security for the case that 3$d$-BSM produces one of the maximally entangled state, {\it e.g.} $\ket{\Phi_{0}}$. In general, the number of possible entangled states that is generated from BSM affects to the sifted key rate. Due to the straightforward relationship, the security proof under the shared single Bell state can be easily generalized to the ideal BSM setup which is able to distinguish all of the nine maximally entangled states.

In the Table, $p(x,y)$ denotes success probability of 3$d$-BSM under the assumption of discriminating single Bell state after Alice and Bob send the states $\ket{x}$ and $\ket{y}$ to Charlie respectively. Table \ref{Table3dMBS} (a) shows the probabilities when the communication sources for the sender's states are prepared perfectly. In Table \ref{Table3dMBS} (b), we assume that Bob's source has misalignment while Alice's is perfect. Bob's source generates $\ket{\bar{\beta}_{0}}$ instead of $\ket{\bar{0}}$. The misaligned state $\ket{\bar{\beta}}$ is defined as $\ket{\bar{\beta}_{0}}=\cos\mu\sin\nu\ket{0}+\sin\mu\sin\nu\ket{1}+\cos\nu\ket{2}$ where $\mu$ and $\nu$ are arbitrary angles. It can be seen that all the probabilities of mismatched-basis case are $1/9$ as in Table \ref{Table3dMBS} (a). In comparison, the distribution of the success probabilities are changed when the basis is misaligned as shown in Table \ref{Table3dMBS} (b). The change allows Alice and Bob to identify the accuracy of communication sources through the analysis of the MBS.

Before we analyze the security,  we defined the explicit form of generated states from uncharacterized sources in detail. The states in the ordinary basis are $\ket{\alpha_{0}}$, $\ket{\alpha_{1}}$, and $\ket{\alpha_{2}}$ for Alice, $\ket{\beta_{0}}$, $\ket{\beta_{1}}$, and $\ket{\beta_{2}}$ for Bob. Subsequently, the states in the bar basis are $\ket{\bar{\alpha}_{0}}$, $\ket{\bar{\alpha}_{1}}$, and $\ket{\bar{\alpha}_{2}}$ for Alice and $\ket{\bar{\beta}_{0}}$, $\ket{\bar{\beta}_{1}}$, and $\ket{\bar{\beta}_{2}}$ for Bob. The relations between two bases are described as
\begin{align}\label{3dUCSrelation}
\ket{\bar{\alpha}_{k}}&=\sum_{j=0}^{2}A_{kj}e^{i\theta_{kj}}\ket{\alpha_{j}}\\
\ket{\bar{\beta}_{k}}&=\sum_{j=0}^{2}B_{kj}e^{i\varphi_{kj}}\ket{\beta_{j}},\nonumber
\end{align}
where $A$, $B$ are non-negative real numbers and $\theta$, $\varphi$ are arbitrary phases. Since Alice and Bob cannot specify the details of generated states, we consider all the possible values of $A$, $B$, $\theta$ and $\varphi$ for our security analysis. The 3-dimensional maximally entangled states are redefined to have arbitrary phases in the generated states. 3-dimensional maximally entangled states with the undetermined phase factors are defined as
\begin{align}\label{3dMEphase}
\ket{\tilde{\Phi}_{3k+l}}=\frac{1}{\sqrt{3}}\sum_{m=0}^{2}\omega^{ml}e^{i(\delta_{m+k}+\xi_{k})}\ket{m+k,m},
\end{align}
where $\delta$ and $\xi$ are phase factors stemmed from experimental setting, and it can be set $\delta_{0}=\xi_{0}=0$.

The secret key rate obtained in Eq.~(\ref{3dQKDrate}) is to be modified for the case of uncharacterized communication source. In the ordinary basis, there are two different types of error $Q_{s}$ and $Q_{p}$. $Q_{s}$ is the state error rate explained in section \ref{sec3dmdi} and $Q_{p}$ is phase error rate which indicates the state error due to the phase factor. If Alice and Bob share the maximally entangled state with additional phase factor $\omega$, for example $\ket{\Phi_{1}}$, then, Alice and Bob identify that there is phase error as it differ from $\ket{\Phi_{0}}$ up to the phase factor in the state. The phase error is comparable with the state error that exists in the bar basis.

Eq.~(\ref{3dQKDrate}) is obtained under the scenario of symmetric attack which is Eve's optimal attack against ideal QKD scheme. The symmetric attack is the strategy to extract the information from the two MUB measurements at both sides equally well \cite{Cerf02,Durt04,Ferenczi12}. Since this attack disturbs the measurement statistics equally for the two different measurement bases, the equivalence of two error rate $Q_{s}=Q_{p}$ can be observed. At the same time, these two errors should be discriminated to analyze the security for QKD with uncharacterized sources. If we split the effect of these two error rates, the secret key rate becomes
\begin{align}\label{3dQKDrateUC}
r\geq\log_{2}3-(Q_{s}+Q_{p})-H(Q_{s})-H(Q_{p}).
\end{align}
So if Alice and Bob calculate $Q_{s}$ and $Q_{p}$, they can evaluate the secret key rate of 3$d$-MDI-QKD with uncharacterized sources.

We analyze the security with uncharacterized sources from the number of generated maximally entangled states using EDP. In the equivalent protocol with uncharacterized sources, Alice and Bob choose measurement basis before generating photon pair. According to their basis choices, they generate the different photon pairs. Alice (Bob) generates $\ket{\psi}_{AC}$ ($\ket{\phi}_{BD}$) if she (he) chooses ordinary basis, and $\ket{\bar{\psi}}_{AC}$ ($\ket{\bar{\phi}}_{BD}$) if not. These state are described as
\begin{align}
\ket{\psi}_{AC}&=\frac{1}{\sqrt{3}}\left( \ket{0, \alpha_{0}}_{AC}+ \ket{1, \alpha_{1}}_{AC}+\ket{2,\alpha_{2}}_{AC} \right)\\
\ket{\bar{\psi}}_{AC}&=\frac{1}{\sqrt{3}}\left( \ket{\bar{0},\bar{\alpha}_{0}}_{AC} +\ket{\bar{1}, \bar{\alpha}_{2}}_{AC} +\ket{\bar{2},\bar{\alpha}_{1}}_{AC} \right)\nonumber\\
\ket{\phi}_{BD}&=\frac{1}{\sqrt{3}}\left( \ket{0, \beta_{0}}_{BD}+ \ket{1, \beta_{1}}_{BD}+\ket{2,\beta_{2}}_{BD} \right)\nonumber\\
\ket{\bar{\phi}}_{BD}&=\frac{1}{\sqrt{3}}\left( \ket{\bar{0},\bar{\beta}_{0}}_{BD} +\ket{\bar{1}, \bar{\beta}_{2}}_{BD} +\ket{\bar{2},\bar{\beta}_{1}}_{BD} \right).\nonumber
\end{align}
Using the entangled states, Alice and Bob proceed 3$d$-MDI-QKD protocol. After several repetition of the procedure, Alice and Bob get statistics of success probabilities of 3$d$-BSM as it is described in Table \ref{Table3dMBS}. The statistics can be used for the security analysis of our protocol.

For the security analysis, we consider the state after the Eve's general attack that can be described as
\begin{align}\label{EveGeattack}
&\hat{U}_{E}\ket{\alpha_{x}}_{C}\ket{\beta_{y}}_{D}\ket{e}_{Ea}\ket{1}_{Z}\\
&=\sqrt{(1-p(x,y))}\ket{\Xi{xy}}_{E}\ket{0}_{Z}+\sqrt{p(x,y)}\ket{\Gamma{xy}}_{E}\ket{1}_{Z}\nonumber
\end{align}
where $\hat{U}_{E}$ is Eve's unitary operation. $\ket{e}_{Ea}$ is Eve's ancillary system and $Z$ denotes the state for the Charlie's message. $\ket{0}_{Z}$ indicates when the BSM fail and $\ket{1}_{Z}$ does BSM succeed. $p(x,y)$ is success probability of 3$d$-BSM when Alice sends $\ket{\alpha_{x}}$ and Bob sends $\ket{\beta_{y}}$ to Charlie. $\ket{\Xi xy}_{E}$ is Eve's final state after eavesdropping when BSM fail, while $\ket{\Gamma xy}_{E}$ is Eve's final state after eavesdropping when BSM succeed. Eve's final state can be expressed as $\ket{\Gamma xy}_{E}=\sum_{n}\gamma_{xy}(n)\ket{n}_{E}$ where $\{\ket{n}\}$ is orthonormal basis of Eve's state. In this expression, $\gamma_{xy}(n)$ is complex number which is obtained from $\gamma_{xy}(n)=\braket{n|\Gamma xy}$. $\gamma_{xy}(n)$ satisfies $\sum_{n}|\gamma_{xy}(n)|^{2}=1$. In these expressions, we meant $x,y\in\{0,1,2,\bar{0},\bar{1},\bar{2}\}$. If we post-select only the case $\ket{1}_{Z}$ when 3$d$-BSM succeed, the state of Alice, Bob, and Eve in the ordinary basis is described as
\begin{align}\label{densityABE}
\hat{\rho}_{ABE}=C \times P\left\{\sum_{x,y=0}\sqrt{p(x,y)}\ket{x,y,\Gamma xy}_{ABE}\right\}
\end{align}
where we use the notation for projector as $P\{\ket{x}\}=\ket{x}\bra{x}$ and $C$ is for the normalization constant. The density operator of Alice and Bob is obtained when Eve's system is traced out.
\begin{align}\label{densityAB}
\hat{\rho}_{AB}&=\text{Tr}_{E}\left[\hat{\rho}_{ABE}\right]\nonumber\\
&=\frac{\sum_{n}P\left\{\sum_{x,y=0}^{2}\sqrt{p(x,y)}\gamma_{xy}(n)\ket{x, y}_{AB}\right\}}{\sum_{x,y=0}^{2}p(x,y)}.
\end{align}

The security proof under the eavesdropping scenario is achieved if Alice and Bob can get $Q_{s}$ and $Q_{p}$. If they get these error rates, they can analyze security of their QKD system from Eq.~(\ref{3dQKDrateUC}). The state error rate $Q_{s}$ is easily obtained from Eq.~(\ref{densityAB}) and Eq.~(\ref{stateerror}). It becomes
\begin{align}\label{stateerrorUC}
Q_{s}=\frac{p(0,1)+p(0,2)+p(1,0)+p(1,2)+p(2,0)+p(2,1)}{\sum_{x,y=0}^{2}p(x,y)}.
\end{align}
From Eq.~(\ref{stateerrorUC}), $Q_{s}$ is calculated from success probabilities of 3$d$-BSM. The phase error rate $Q_{p}$ is obtained by
\begin{align}
Q_{p}=&\bra{\widetilde{\Phi}_{1}}\hat{\rho}\ket{\widetilde{\Phi}_{1}} +\bra{\widetilde{\Phi}_{2}}\hat{\rho}\ket{\widetilde{\Phi}_{2}} +\bra{\widetilde{\Phi}_{4}}\hat{\rho}\ket{\widetilde{\Phi}_{4}} \nonumber\\
&+\bra{\widetilde{\Phi}_{5}}\hat{\rho}\ket{\widetilde{\Phi}_{5}}+\bra{\widetilde{\Phi}_{7}}\hat{\rho}\ket{\widetilde{\Phi}_{7}} +\bra{\widetilde{\Phi}_{8}}\hat{\rho}\ket{\widetilde{\Phi}_{8}}. \nonumber
\end{align}
As the last 4 terms are included in the state error rate, the equation is simplified by the  inequality,
\begin{align}\label{PBwithBell}
Q_{p} \leq & \bra{\widetilde{\Phi}_{1}}\hat{\rho}\ket{\widetilde{\Phi}_{1}} +\bra{\widetilde{\Phi}_{2}}\hat{\rho}\ket{\widetilde{\Phi}_{2}} +Q_{s}.
\end{align}
$\bra{\widetilde{\Phi}_{1}}\hat{\rho}\ket{\widetilde{\Phi}_{1}} $ and $\bra{\widetilde{\Phi}_{2}}\hat{\rho}\ket{\widetilde{\Phi}_{2}}$ include phase factors $\delta$ and $\xi$ which are defined in Eq.~(\ref{3dMEphase}). In the circumstance, Alice and Bob do not know the details of the phase factors since the communication sources are uncharacterized. For the true upper bound, one should consider the largest values of $\bra{\widetilde{\Phi}_{1}}\hat{\rho}\ket{\widetilde{\Phi}_{1}}$ and $\bra{\widetilde{\Phi}_{2}}\hat{\rho}\ket{\widetilde{\Phi}_{2}}$. With the maximum, the new upper bound of phase error rate can be found as
\begin{align}\label{bPEupperbound}
Q_{p}\leq& \frac{2}{3}\frac{\sum_{n}\left| \sum_{k=0}^{2}\sqrt{p(k,k)}e^{i\zeta_{k}}\gamma_{kk}(n) \right|^{2}}{\sum_{x,y=0}^{2}p(x,y)} +Q_{s}
\end{align}
where $\zeta$ is the phase which makes the first term maximum. There are $\gamma$ and $\zeta$ factors comes from Eve's states. Alice and Bob can not determine these factors. What Alice and Bob want to get is the upper bound of phase error rate which can be calculated from the success probabilities of 3$d$-BSM only. We formulate the new upper bound of phase error rate as
\begin{align}\label{phasebound}
Q_{p}\leq \varepsilon+Q_{s},
\end{align}
where $\varepsilon$ is the factor calculated from the success probabilities that is obtained from Eq.~(\ref{3dUCSrelation}) and Eq.~(\ref{EveGeattack}). It is defined as
\begin{align}\label{varepsilon}
\varepsilon=\max_{A,B}\left[f_{01}(A,B),f_{20}(A,B)\right],
\end{align}
where max means finding maximum value for the coefficients $A$, $B$ defined in Eq.~(\ref{3dUCSrelation}).
Details of the calculation is described in Appendix. The function $f$ is defined as
\begin{widetext}
\begin{align}\label{upperfunction}
f_{xy}(A,B)=
\left\{\begin{matrix}
1-Q_{s} & \text{if } A_{xm}B_{ym}= 0, \quad \forall m\in\{0,1,2\}\\
\min[S_{xy}(0), S_{xy}(1), S_{xy}(2)] & \text{otherwise.}
\end{matrix}\right.
\end{align}
$S_{xy}(m)$ is the upper bound function of the first term in Eq.~(\ref{bPEupperbound}) defined as
\begin{align}\label{PBfunction}
S_{xy}(m)=\frac{2}{3A_{xm}^{2}B_{ym}^{2}\sum_{i,j=0}^{2}p(i,j)}\left\{\sqrt{\left[\sqrt{p(\bar{x},\bar{y})}+\sum_{i\neq j}A_{xi}B_{yj}\sqrt{p(i,j)}\right]^{2}+2\prod_{k=1}^{2} D_{xym}(k)}+\sum_{k=1}^{2}D_{xym}(k)\right\}^{2},
\end{align}
where $D_{xym}(k)$ denotes the expression
\begin{align}
D_{xym}(k)=|A_{xm}B_{ym}-A_{x(m+k)}B_{y(m+k)}|\sqrt{p(m+k,m+k)}.\nonumber
\end{align}
\end{widetext}
The function $f$ is formed to find the smallest value of $S_{xy}(m)$ for $m$ in order to obtain the tight upper bound of the first term of Eq.~(\ref{bPEupperbound}). $S_{xy}(m)$ exists only when $A_{xm}B_{ym}\neq0$, because of the factor $1/A_{xm}B_{ym}$. If $A_{xm}B_{ym}=0$ for all $m\in\{0,1,2\}$, no upper bound function $S_{xy}(m)$ can be determined. In this case, Alice and Bob conclude that the states generated from the uncharacterized sources are not suitable for QKD. The function $f$ has the form that the phase error rate becomes one in the case.

$\varepsilon$ is defined to identify the relations between $\ket{\bar{\alpha}_{0}}$ and $\ket{\bar{\beta}_{1}}$ as well as $\ket{\bar{\alpha}_{2}}$ and $\ket{\bar{\beta}_{0}}$ in Eq.~(\ref{varepsilon}). The success probabilities of 3$d$-BSM for these states are $p(\bar{0},\bar{1})=0$ and $p(\bar{2},\bar{0})=0$ in the ideal 3$d$-MDI-QKD. If communication sources are ideal, $\varepsilon$ is zero. Then the secret key rate of $d$-MDI-QKD with uncharacterized sources in Eq.~(\ref{3dQKDrateUC}) becomes same with the secret key rate of ideal 3$d$-MDI-QKD in Eq.~(\ref{3dQKDrate}). If the states generated from uncharacterized sources are not the ideal states, then the factor $\varepsilon$ becomes non-zero. In the case, the maximization of $\varepsilon$ is required in order to obtain the optimized secret key rate.

In order to find the maximum value of the factor $\varepsilon$, Alice and Bob need to consider the constraints about the coefficients $A$ and $B$. The constraints are obtained from MBS under the Eve's attack, modeled in Eq.~(\ref{EveGeattack}). In the case that Alice and Bob send the single photon states $\ket{\bar{\alpha}_{x}}$ and $\ket{\beta_{y}}$ to Charlie respectively, the success probability of 3$d$-BSM can be obtained as
\begin{align}\nonumber
p(\bar{x},y)=|_{Z}\bra{1}_{E}\bra{\Gamma \bar{x}y}\hat{U}_{E}\ket{\bar{\alpha}_{x}}_{A}\ket{\beta_{y}}_{B}|^{2}
\end{align}
and if we substitute $\ket{\bar{\alpha}_{x}}$ in Eq.~(\ref{3dUCSrelation}), we have
\begin{align}\label{constraintequation}
p(\bar{x},y)=|_{Z}\bra{1}_{E}\bra{\Gamma \bar{x}y}\hat{U}_{E}\left[\sum_{i=0}^{2}A_{xi}e^{\theta_{xi}}\ket{\alpha_{i}}_{A}\right]\ket{\beta_{y}}_{B}|^{2}.
\end{align}
Since Alice and Bob cannot determine the parameters in the Eve's side, $\ket{\Gamma xy}_{E}$ and $\theta$, they need to consider the largest and the smallest values of undetermined factors for the constraints. Then, the constraints for Alice's coefficient can be obtained from Eq.~(\ref{constraintequation}),
\begin{gather}
-2\sum_{l=0}^{2}\left[A_{jl}A_{j(l+1)}\sqrt{p(l,i)p(l+1,i)}\right]\nonumber\\
\leq p(\bar{j},i) -\sum_{l=0}^{2}A_{jl}^{2}p(l,i)\leq\label{constraintA}\\
2 \sum_{l=0}^{2}\left[A_{jl}A_{j(l+1)}\sqrt{p(l,i)p(l+1,i)}\right]\nonumber
\end{gather}
where $i\in\{0,1,2\}$ and $j\in\{0,2\}$. Similarly, the constraints for Bob's coefficient can be obtained from the success probability of 3$d$-BSM when Alice and Bob send $\ket{\alpha_{x}}$ and $\ket{\bar{\beta}_{y}}$ to Charlie respectively. The constraints for Bob's states are
\begin{gather}
-2\sum_{l=0}^{2}\left[B_{kl}B_{k(l+1)}\sqrt{p(i,l)p(i,l+1)}\right]\nonumber\\
\leq p(i,\bar{k})-\sum_{l=0}^{2}B_{kl}^{2}p(i,l)\leq\label{constraintB}\\
2\sum_{l=0}^{2}\left[B_{kl}B_{k(l+1)}\sqrt{p(i,l)p(i,l+1)}\right]\nonumber
\end{gather}
where $i\in\{0,1,2\}$ and $k\in\{1,0\}$. Now, Alice and Bob have six constraints at each side. Using the constraints, Alice and Bob are able to calculate the upper bound of phase error rate with Eq.~(\ref{phasebound}). Finally the secret key rate of 3$d$-MDI-QKD with uncharacterized source can also be obtained from the success probabilities of 3$d$-BSM.

\section{Simulation}\label{secsimul}

\begin{figure*}[t]
\begin{subfigure}[]{\includegraphics[width=0.31\textwidth]{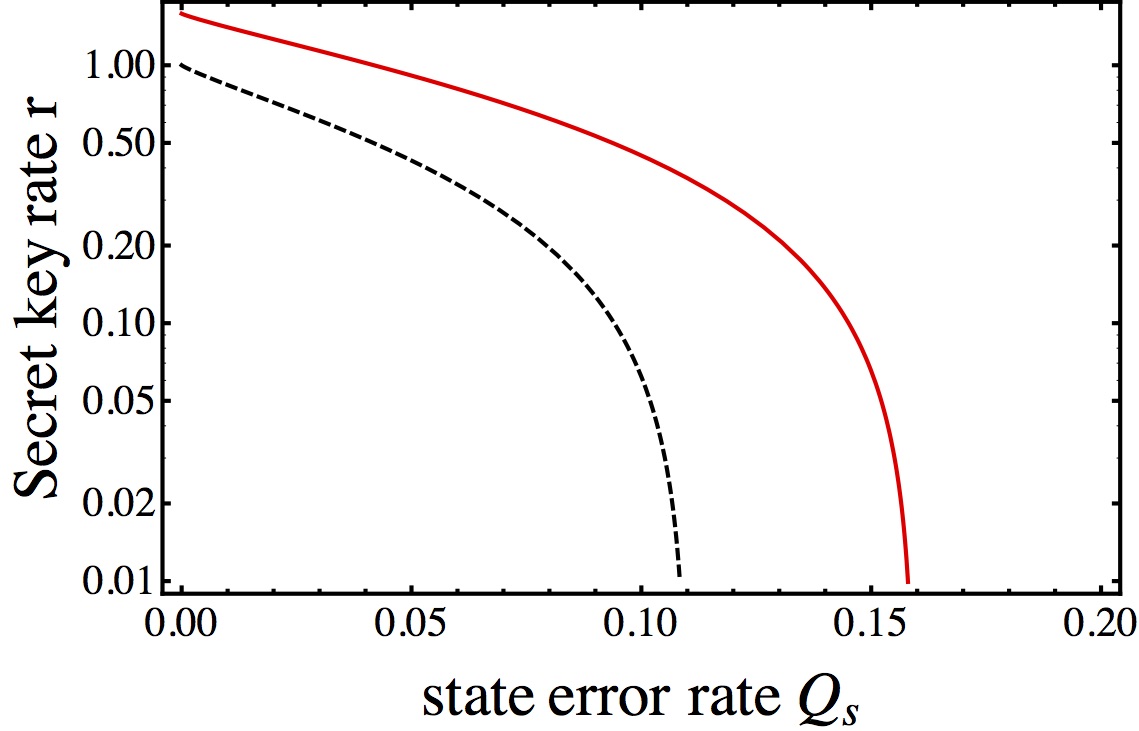}}\end{subfigure}
\begin{subfigure}[]{\includegraphics[width=0.31\textwidth]{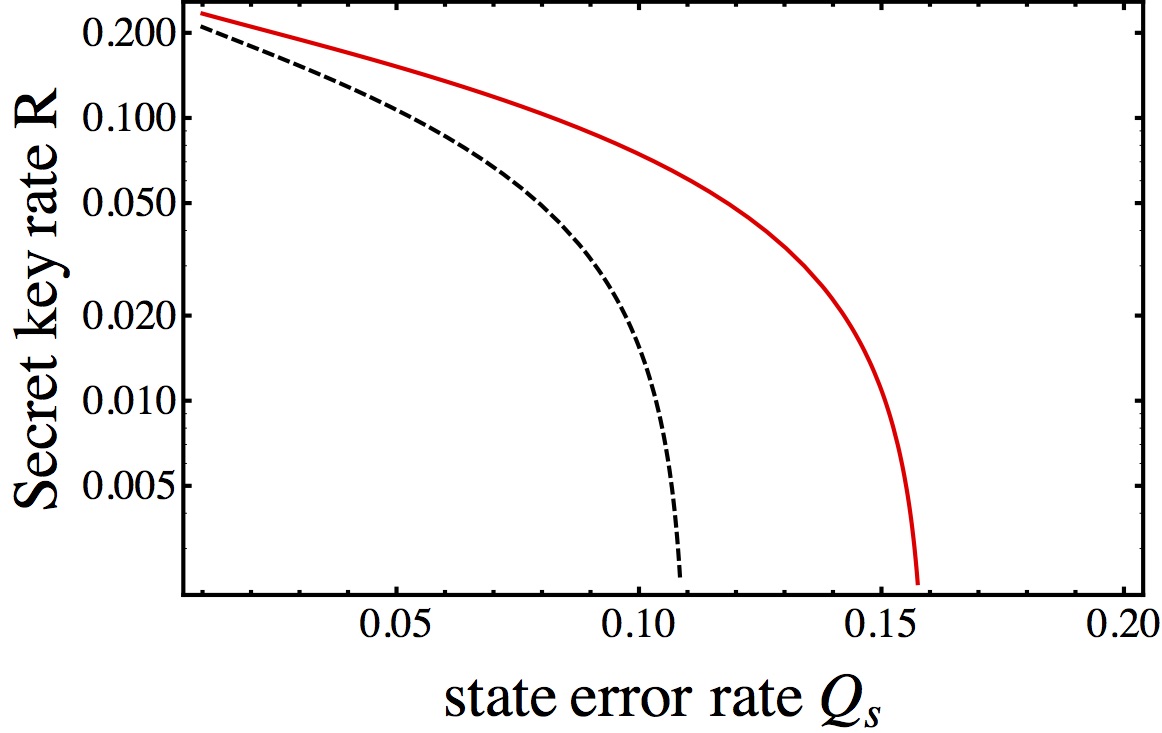}}\end{subfigure}
\begin{subfigure}[]{\includegraphics[width=0.31\textwidth]{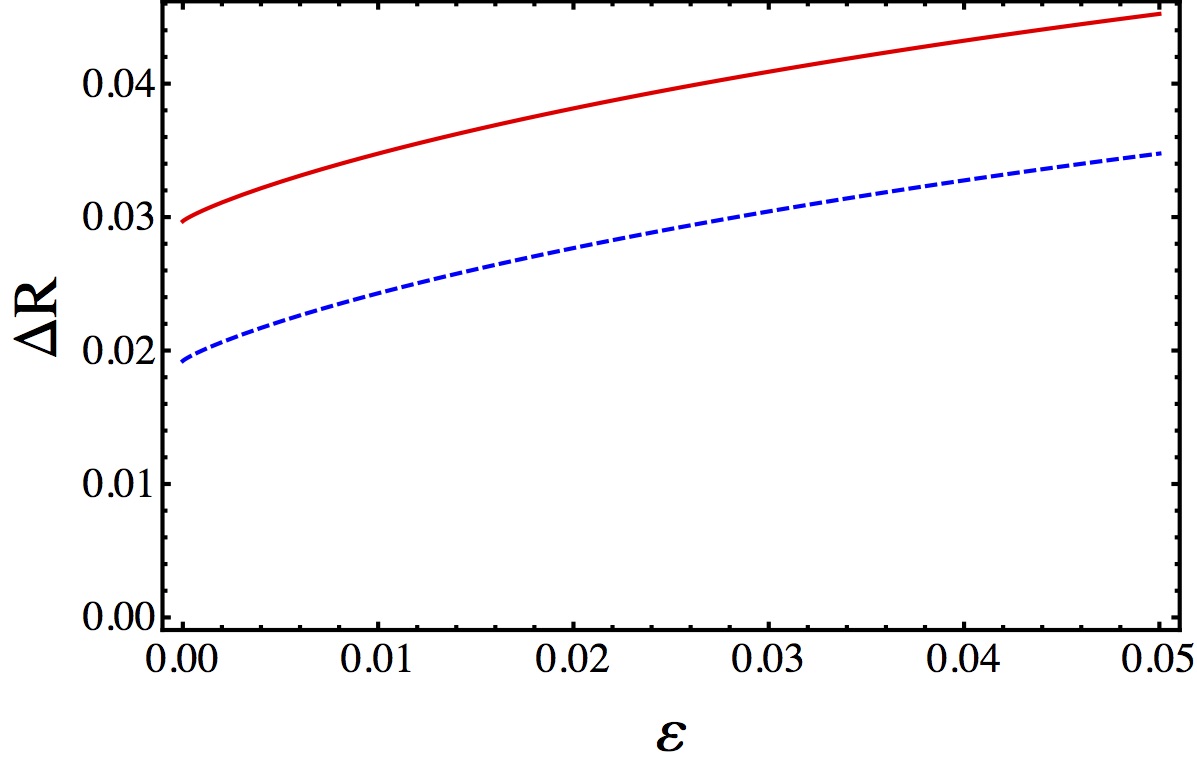}}\end{subfigure}
\caption{(a) The secret key rate per sifted key $r$ of original MDI-QKD (black, dashed line) and 3$d$-MDI-QKD (red line). (b) Secret key rate $R$ per total signal of original qubit MDI-QKD (black, dashed line) and 3$d$-MDI-QKD (red line). We assumed 3$d$-BSM is able to discriminate three maximally entangled states among nine, and qubit BSM can discriminate two Bell states among four. (c) The difference between the secret key rate $R$ of original MDI-QKD and 3$d$-MDI-QKD. Both secret key rates are obtained under the assumption of uncharacterized sources. $\varepsilon$ is the factor which is defined in Eq.~(\ref{phasebound}). Blue, dashed line shows difference between the secret key rates when the state error rate is fixed as $Q_{s}=0.01$, red line shows the difference when $Q_{s}=0.05$.}\label{figIdsimul}
\end{figure*}
\begin{figure*}[t]
\begin{subfigure}[]{\includegraphics[width=0.31\textwidth]{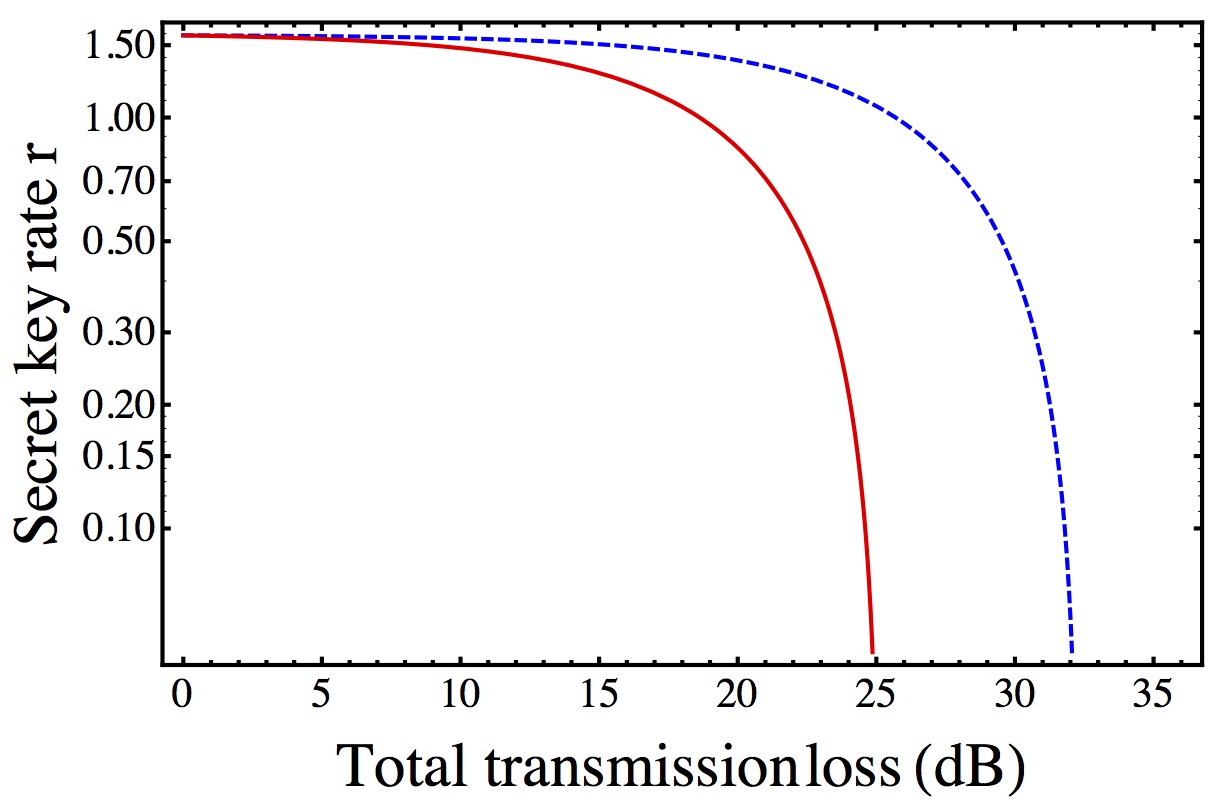}}\end{subfigure}
\begin{subfigure}[]{\includegraphics[width=0.31\textwidth]{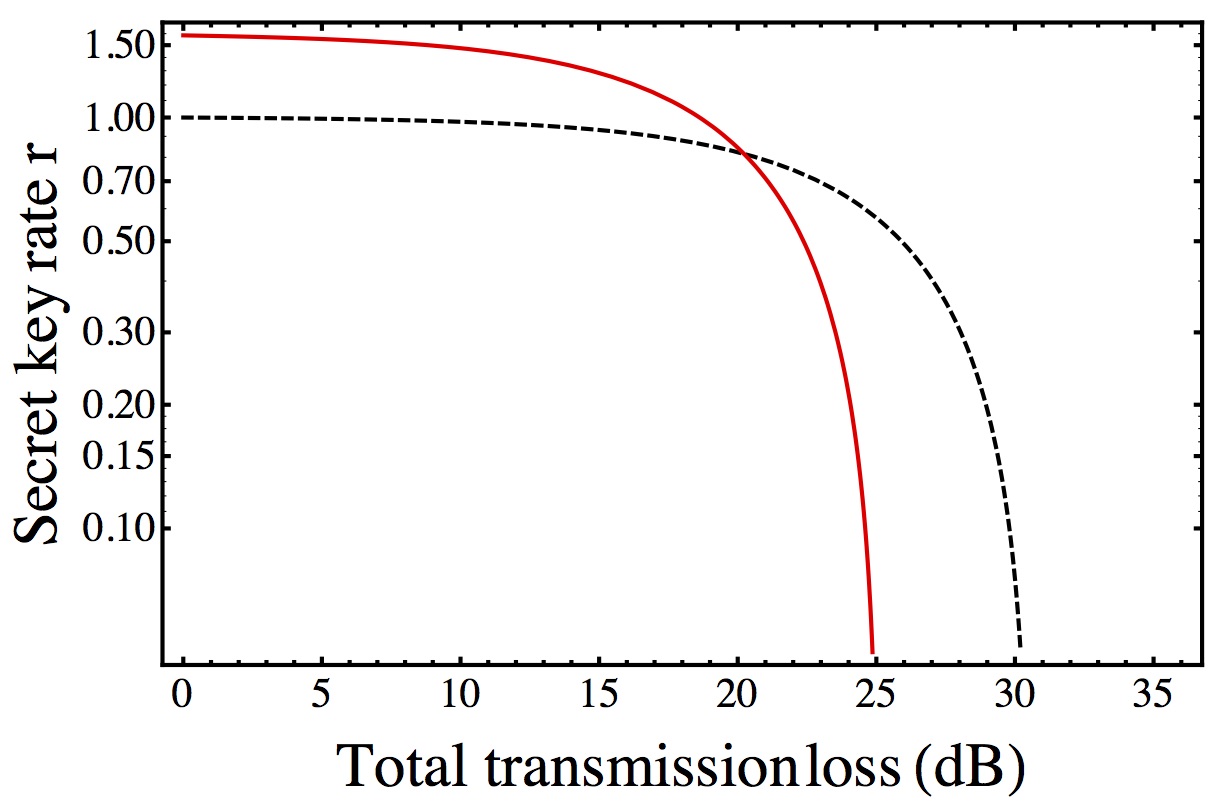}}\end{subfigure}
\begin{subfigure}[]{\includegraphics[width=0.31\textwidth]{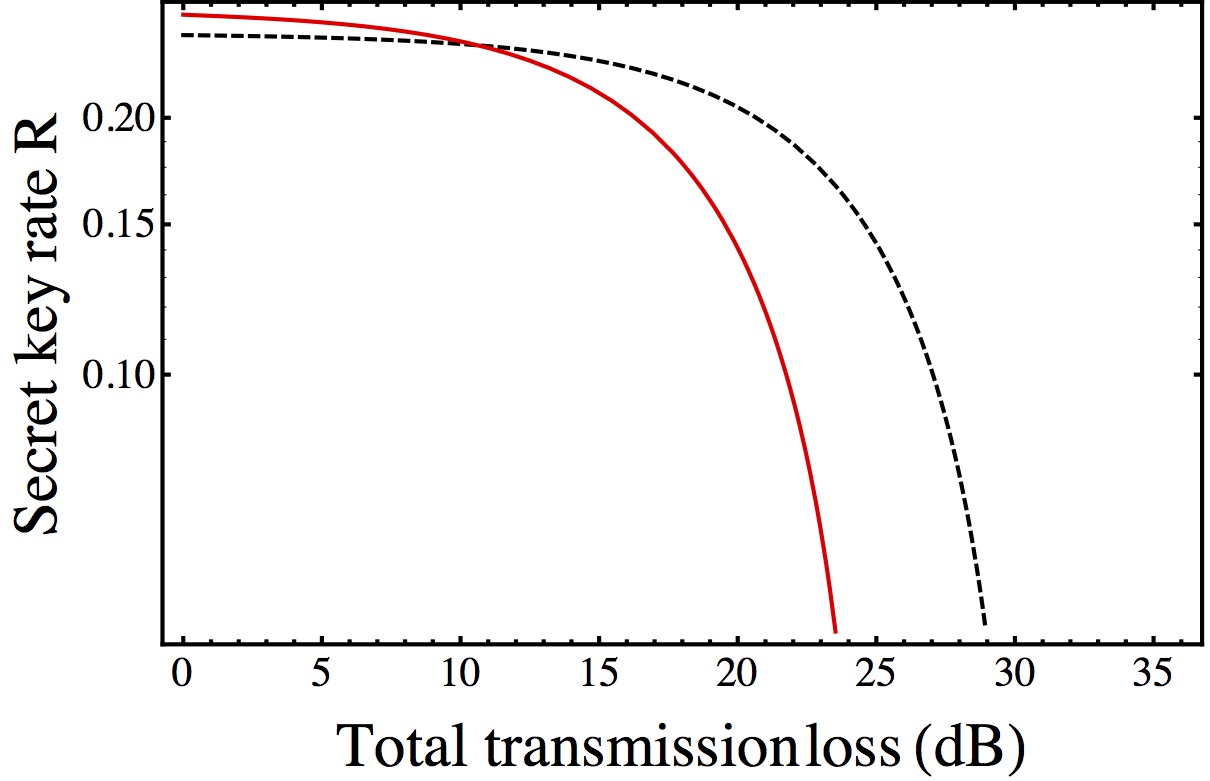}}\end{subfigure}
\caption{The secret key rate of MDI-QKD vs. total transmission loss of optical channel. The dark count rate of single photon detector is assumed as $10^{-5}$ per pulse. We assume there is no Eve. (a) The secret key rate $r$ per sifted key of 3$d$-MDI-QKD (blue, dashed line) and 3$d$-MDI-QKD with uncharacterized sources (red line). (b) The secret key rate $r$ per sifted key of original MDI-QKD (black, dashed line) and 3$d$-MDI-QKD (red line). Both secret key rates are obtained under the assumption of uncharacterized sources. The intersection point of the secret key rates appears at the transmission loss 20dB approximately. (c) The secret key rate $R$ per total signal of original MDI-QKD (black, dashed line) and 3$d$-MDI-QKD (red line) under the assumption of uncharacterized sources. We assume that qubit BSM discriminates two Bell states among four and 3$d$-BSM does three maximally entangled states among nine. The intersection point appears at transmission loss 10.5dB approximately.}\label{figResimul}
\end{figure*}

In this section, we present the results for the secret key rate simulation of 3$d$-MDI-QKD and 3$d$-MDI-QKD with uncharacterized sources. In the simulation, we consider the communication scenario that is realized by single photon source only. First of all, we discuss the ideal 3$d$-MDI-QKD with uncharacterized sources. Since there is no Eve and the devices are perfectly implemented, the statistics of success probabilities in 3$d$-BSM becomes same as it is shown in Table \ref{Table3dMBS} (a). The state error rate and the phase error rate are both zero within the constraints in Eq.~(\ref{constraintA}) and in Eq.~(\ref{constraintB}). Therefore, in the ideal case, the secret key rate of 3$d$-MDI-QKD is identical to that of the protocol with uncharacterized sources.

Now, we compare the secret key rate $r$ of original qubit MDI-QKD \cite{Lo12} and 3$d$-MDI-QKD. In our simulation, $r$ denotes the secret key rate per sifted key. The secret key rate per sifted key of original qubit MDI-QKD is $r_{2}=1-2H(Q_{b})$, where $Q_{b}$ is quantum bit error rate (QBER). The secret key rate per sifted key of 3$d$-MDI-QKD is given in Eq.~(\ref{3dQKDrate}). Fig.~\ref{figIdsimul} (a) shows these two secret key rates. Red line denotes the secret key rate of 3$d$-MDI-QKD and black dashed line denotes that of original MDI-QKD. As it is seen already \cite{Cerf02, Durt04, Ferenczi12, Sheridan10}, the secret key rate of 3$d$-MDI-QKD is higher than that of original MDI-QKD at the same error rate.

The original MDI-QKD has lower secret key rate than BB84 since it has lower sifted key rate. We consider the sifted key rate in the simulation and show that 3$d$-MDI-QKD has higher secret key rate than original one even in the realistic experiments. Fig.~\ref{figIdsimul} (b) shows the secret key rate $R$ of original MDI-QKD and 3$d$-MDI-QKD per total signal. $R$ denotes the secret key rate per total signal. The total signal includes the trials that Alice and Bob can not generate the maximally entangled state and mismatched bases are used in the measurements.

The secret key rate $R$ can be obtained from $R=r\times$(sifted key rate). The sifted key rate is defined as (sifted key rate)=(the probability Alice and Bob choose same basis)$\times$(the success probability of BSM). Since the original MDI-QKD and 3$d$-MDI-QKD use two MUBs, the first probability is same with $1/2$. The success probability of BSM is $1/2$ since qubit BSM setup can discriminate two Bell states among four Bell states \cite{Lutkenhaus99}. There is no proposed 3$d$-BSM setup with passive linear optical elements yet. In the simulation, we assume that 3$d$-BSM discriminate three maximally entangled states among nine. In that case, the success probability of 3$d$-BSM is $1/3$. The secret key rate of original MDI-QKD $R_{2}$ is obtained from $R_{2}=\frac{1}{4}r_{2}$ and that of 3$d$-MDI-QKD is $R_{3}=\frac{1}{6}r_{3}$. In the Fig.~\ref{figIdsimul} (b), the red line denotes $R_{3}$ and the black, dashed line shows $R_{2}$. Since the success probability of 3$d$-BSM is lower than that of qubit BSM, the difference between two key rates is decreased, but still 3$d$-MDI-QKD always has higher key rate.

Fig.~\ref{figIdsimul} (c) shows the difference between the two key rates $\Delta R=R_{3}-R_{2}$ against $\varepsilon$ which is defined in Eq.~(\ref{phasebound}). The factor $\varepsilon$ is related with suitability of sources for QKD. The red line and the blue, dashed line denote the case $Q_{s}=0.01$ and $Q_{s}=0.05$ respectively. $\Delta R$ becomes large when $\varepsilon$ goes large regardless the state error rate.

Here, we simulate the secret key rate with the change of realistic experimental factors. We consider the two experimental factors, transmission efficiency of photonic channels ($\eta$) and dark count rate of single photon detectors (SPDs) $d$. There is transmission loss when the photon passes through optical fiber or atmosphere, so the transmission efficiency is approximately proportional to the distance which QKD can be achieved. For SPD, since the SPD is very sensitive device, it is possible to be clicked even if no photon enters in SPD. This click produce the dark count in the device. The dark count rate is assumed as $10^{-5}$ per pulse in the our simulation.

Qubit BSM consists of four SPDs \cite{Lutkenhaus99} in order to discriminate polarization Bell states of the photons. Contrarily, as it is already mentioned, there is no realistic 3$d$-BSM setup with passive linear optical elements proposed yet. In our simulation, we assume that 3$d$-BSM consists of six SPDs to measure two photons and each photon is encoded in three orthonormal states. If there is no Eve, the success probabilities of 3$d$-BSM is described as
\begin{align}
&p(i,i)=\label{probsimulsame}\\
&\frac{\eta^{2}(1-d)^{4}}{3} +2 \eta (1-\eta) d (1-d)^{4}+3 (1-\eta)^{2} d^{2} (1-d)^{4} \nonumber \\
&p(i,j)=2 \eta (1-\eta) d (1-d)^{4}+3 (1-\eta)^{2} d^{2} (1-d)^{4} \label{probsimuldiff}
\end{align}
where $i,j\in\{0,1,2\}$, $i\neq j$ and 3$d$-BSM discriminates one of the maximally entangled state $\ket{\Phi_{0}}$ only. In that case, all the probabilities of mismatched basis cases are same, so that  $p(i,\bar{j})=p(j,\bar{i})=p(\bar{i},j)$ where $i,j\in\{0,1,2\}$. In Eq.~(\ref{probsimulsame}), the first term describes the situation that two photons trigger off SPDs with $1/3$ success probability. The second term denotes the situation that one photon arrive at SPD and the other photon is lost. In that case, the second SPD is clicked due to the dark count of the detector. The final term denotes the situation that two photons are lost but two SPDs are clicked because of dark counts \cite{Yin14}. With these probabilities, the state error rate and the phase error rate can be calculated and the secret key rates are plotted in Fig.~\ref{figResimul} with respect to ($1-\eta$).

Fig.~\ref{figResimul} (a) shows the secret key rate $r$ of 3$d$-MDI-QKD (blue, dashed line) and 3$d$-MDI-QKD with uncharacterized sources (red line) vs. transmission loss. $\varepsilon$ is not zero because of transmission loss in Eq.~(\ref{probsimuldiff}). Since non-zero value of $\varepsilon$ is considered as imperfection of communication sources in the security analysis under the assumption of uncharacterized source, 3$d$-MDI-QKD with uncharacterized sources has lower the secret key rate than 3$d$-MDI-QKD even if communication sources generate the exact states.

The secret key rate $r$ of original MDI-QKD (black, dashed line) and 3$d$-MDI-QKD (red line) are shown in Fig.~\ref{figResimul} (b). Both secret key rates are calculated under the assumption of uncharacterized sources \cite{Yin14}. 3$d$-MDI-QKD has higher secret key rate than original MDI-QKD only when transmission loss is low. This effect comes from increased dark counts since 3$d$-BSM has more SPDs than qubit BSM. The cross point appears at the transmission loss 20dB approximately.

Fig.~\ref{figResimul} (c) shows the secret key rate $R$ of original MDI-QKD (black, dashed line) and 3$d$-MDI-QKD (red line). These secret key rates are calculated from the multiplication between sifted key rate and the secret key rates plotted $r$ shown in Fig.~\ref{figResimul} (b). Because of low success probability of 3$d$-BSM, the difference between two key rates is decreased. The cross point appears at the transmission loss 10.5dB approximately.

\section{Conclusion}\label{secconclusion}

In this paper, we investigated security of 3$d$-MDI-QKD and that of 3$d$-MDI-QKD with uncharacterized sources. We assumed that 3$d$-BSM is able to discriminate three maximally entangled states among nine in the simulation. We showed that 3$d$-MDI-QKD has higher secret key rate than original qubit MDI-QKD at the same error rate even if we consider sifted key rate. The endurance of 3$d$-MDI-QKD against low accuracy of the communication sources is better than that of original MDI-QKD as shown in Fig.~\ref{figIdsimul} (c). In the simulation with the realistic experimental parameters, 3$d$-MDI-QKD has higher key rates only when the transmission loss is low. Since the transmission loss proportional to length of optical fiber, this shows that 3$d$-MDI-QKD has higher secret key rate only when it is implemented for the short distance communication.

\section{Acknowledgement}
\indent This work was done with support of ICT R\&D program of MSIP/IITP (No.2014-044-014-002), National Research Foundation(NRF) grant (No.NRF-2013R1A1A2010537), and National Research Council of Science and Technology(NST) grant (No. CAP-15-08-KRISS).
\numberwithin{equation}{section}

\begin{appendix}
\section{Calculation of $\varepsilon$ in 3$d$-MDI-QKD with uncharacterized sources}\label{app3dMBSMDI-QKD}

In appendix, we show the details of calculation to obtain Eq.~(\ref{PBfunction}). The upper bound of phase error rate is
\begin{align}
Q_{p}\leq\bra{\widetilde{\Phi}_{1}}\hat{\rho}\ket{\widetilde{\Phi}_{1}} +\bra{\widetilde{\Phi}_{2}}\hat{\rho}\ket{\widetilde{\Phi}_{2}}+Q_{s}
\end{align}
from Eq.~(\ref{PBwithBell}). The first two terms can be calculated from density operator of Alice and Bob
\begin{align}
\bra{\widetilde{\Phi}_{1}}\hat{\rho}\ket{\widetilde{\Phi}_{1}}&= \frac{1}{3}\frac{\sum_{n}|\sum_{k=0}^{2}\sqrt{p(k,k)}\omega^{k}e^{i(\delta_{k}+\xi_{k})}\gamma_{kk}(n)|^{2}}{\sum_{x,y=0}^{2}p(x,y)},\nonumber\\
\bra{\widetilde{\Phi}_{2}}\hat{\rho}\ket{\widetilde{\Phi}_{2}}&= \frac{1}{3}\frac{\sum_{n}|\sum_{k=0}^{2}\sqrt{p(k,k)}\omega^{2k}e^{i(\delta_{k}+\xi_{k})}\gamma_{kk}(n)|^{2}}{\sum_{x,y=0}^{2}p(x,y)},\nonumber
\end{align}
where $\delta_{0}=\xi_{0}=0$. Since Alice and Bob can not determine the exact phases $\delta$ and $\xi$, they should consider the maximum values of these two equation in order to cover all the possible phase factors. Then the upper bound of the phase error rate becomes
\begin{align}
Q_{p}\le \frac{2}{3}\frac{\sum_{n}|\sum_{k=0}^{2}\sqrt{p(k,k)}e^{i\zeta_{k}}\gamma_{kk}(n)|^{2}}{\sum_{x,y=0}^{2}p(x,y)}+Q_{s}.\nonumber
\end{align}
where $\zeta$ is the phase which makes maximum the first term. There are $\gamma$ and $\zeta$ factors which Alice and Bob can not determine. Alice and Bob need the upper bound of the phase error rate which is described with the success probabilities of 3$d$-BSM only. The phase error rate in the ordinary basis must be related with probabilities in bar basis. Therefore, we consider the caset that Alice and Bob sends $\ket{\bar{\alpha}_{x}}$ and $\ket{\bar{\beta}_{y}}$ to Charlie respectively. After post-selection of successful 3$d$-BSM case($\ket{1}_{Z}$), Eq.~(\ref{EveGeattack}) becomes
\begin{align}\label{barattack}
\hat{U}_{E}\ket{\bar{\alpha}_{x}}_{C}\ket{\bar{\beta}_{y}}\ket{e}_{Ea}&=\sqrt{p(\bar{x},\bar{y})}\ket{\Gamma \bar{x}\bar{y}}_{E}\nonumber\\
&=\sqrt{p(\bar{x},\bar{y})}\sum_{n}\gamma_{\bar{x}\bar{y}}(n)\ket{n}_{E}.
\end{align}
\begin{widetext}
The left-hand-side of Eq.~(\ref{barattack}) is described in the ordinary basis from the relations between two bases(Eq.~(\ref{3dUCSrelation})),
\begin{align}
\hat{U}_{E}\ket{\bar{\alpha}_{x}}_{C}\ket{\bar{\beta}_{y}}_{D}\ket{e}_{Ea}&=\hat{U}_{E}\sum_{i,j=0}^{2}A_{xi}B_{yj}e^{i(\theta_{xi}+\varphi_{yj})}\ket{\alpha_{i}}_{C}\ket{\beta_{j}}_{D}\ket{e}_{Ea}\nonumber\\
&=\sum_{i,j=0}^{2}A_{xi}B_{yj}\sqrt{p(i,j)}e^{i(\theta_{xi}+\varphi_{yj})}\ket{\Gamma ij}_{E}\nonumber\\
&=\sum_{n}\sum_{i,j=0}^{2}A_{xi}B_{yj}\sqrt{p(i,j)}e^{i(\theta_{xi}+\varphi_{yj})}\gamma_{ij}(n)\ket{n}_{E}.\nonumber
\end{align}
With this equation, Eq.~(\ref{barattack}) becomes
\begin{align}
\sum_{n}\sum_{i,j=0}^{2}A_{xi}B_{yj}\sqrt{p(i,j)}e^{i(\theta_{xi}+\varphi_{yj})}\gamma_{ij}(n)\ket{n}_{E}=\sqrt{p(\bar{x},\bar{y})}\sum_{n}\gamma_{\bar{x}\bar{y}}(n)\ket{n}_{E}.
\end{align}
We rearrange the equation to obtain the upper bound of the phase error rate,
\begin{align}
&\sum_{n}\sum_{i=0}^{2}A_{xi}B_{yi}\sqrt{p(i,i)}e^{i(\theta_{xi}+\varphi_{yi})}\gamma_{ii}(n)\ket{n}_{E}\nonumber\\
&=\sum_{n}\left[\sqrt{p(\bar{x},\bar{y})}\gamma_{\bar{x}\bar{y}}(n) -\sum_{i \neq j}^{2}A_{xi}B_{yj}\sqrt{p(i,j)}e^{i(\theta_{xi}+\varphi_{yj})}\gamma_{ij}(n)\right]\ket{n}_{E}.
\end{align}
Doing absolute square on both sides,
\begin{align}
\sum_{n}\left|\sum_{i=0}^{2}A_{xi}B_{yi}\sqrt{p(i,i)}e^{i(\theta_{xi}+\varphi_{yi})}\gamma_{ii}(n)\right|^{2}&=p(\bar{x},\bar{y})+\sum_{n}\left|\sum_{i\neq j}^{2}A_{xi}B_{yj}\sqrt{p(i,j)}e^{i(\theta_{xi}+\varphi_{yj})}\gamma_{ij}(n)\right|^{2}\\
&\quad-\sum_{n}\sqrt{p(\bar{x},\bar{y})}\gamma_{\bar{x}\bar{y}}^{\ast}(n)\sum_{i\neq j}^{2}A_{xi}B_{yj}\sqrt{p(i,j)}e^{i(\theta_{xi}+\varphi_{yj})}\gamma_{ij}(n) \nonumber\\
&\quad -\sum_{n}\sqrt{p(\bar{x},\bar{y})}\gamma_{\bar{x}\bar{y}}(n)\sum_{i\neq j}^{2}A_{xi}B_{yj}\sqrt{p(i,j)}e^{-i(\theta_{xi}+\varphi_{yj})}\gamma_{ij}^{\ast}(n),\nonumber
\end{align}
\noindent since $\sum_{n}|\gamma_{xy}(n)|^{2}=1$. Considering the phase which makes right-hand-side maximum,
\begin{align}\label{intercal}
\sum_{n}\left|\sum_{i=0}^{2}A_{xi}B_{yi}\sqrt{p(i,i)}e^{i(\theta_{xi}+\varphi_{yi})}\gamma_{ii}(n)\right|^{2}&\leq \left[\sqrt{p(\bar{x},\bar{y})}+\sum_{i\neq j}^{2}A_{xi}B_{yj}\sqrt{p(i,j)}\right]^{2}.
\end{align}
The inequality is satisfied for all phase $\theta$ and $\varphi$, so
\begin{align}\label{nophaseineq}
&\sum_{n}\left|\sum_{i=0}^{2}A_{xi}B_{yi}\sqrt{p(i,i)}e^{i\zeta_{i}}\gamma_{ii}(n)\right|^{2}\leq\left[\sqrt{p(\bar{x},\bar{y})}+\sum_{i\neq j}^{2}A_{xi}B_{yj}\sqrt{p(i,j)}\right]^{2}
\end{align}
is also satisfied, where $\zeta$ is the phase makes left-hand-side maximum. By using triangular inequality, the left-hand-side of the inequality becomes
\begin{align}\label{triangleineq}
&\sum_{n}\left|\sum_{i=0}^{2}A_{xi}B_{yi}\sqrt{p(i,i)}e^{i\zeta_{i}}\gamma_{ii}(n)\right|^{2} \\
&\geq \sum_{n}\left[A_{xm}B_{ym}\left|\sum_{i=0}^{2}\sqrt{p(i,i)}e^{i\zeta_{i}}\gamma_{ii}(n)\right|-\sum_{i=0}^{2}|A_{xm}B_{ym}-A_{xi}B_{yi}|\sqrt{p(i,i)}|e^{i\zeta_{i}}\gamma_{ii}(n)|\right]^{2}\nonumber\\
&=\sum_{n}\left[A_{xm}^{2}B_{ym}^{2}\left|\sum_{i=0}^{2}\sqrt{p(i,i)}e^{i\zeta_{i}}\gamma_{ii}(n)\right|^{2}+\sum_{i=0}^{2}|A_{xm}B_{ym}-A_{xi}B_{yi}|^{2}p(i,i)|e^{i\zeta_{i}}\gamma_{ii}(n)|^{2}\right.\nonumber\\
&\quad -\left.2A_{xm}B_{ym}\left|\sum_{i=0}^{2}\sqrt{p(i,i)}e^{i\zeta_{i}}\gamma_{ii}(n)\right|\sum_{k=1}^{2}D_{xym}(k)|e^{i\zeta_{m+k}}\gamma_{(m+k)(m+k)}(n)|+2\prod_{k=1}^{2}D_{xym}(k)|e^{i\zeta_{m+k}}\gamma_{(m+k)(m+k)}|\right]\nonumber
\end{align}
where we omit (mod 3) in subscription, $m\in\{0,1,2\}$, and $D_{xym}(k)$ is the expression defined as
\begin{align}
D_{xym}(k)=|A_{xm}B_{ym}-A_{x(m+k)}B_{y(m+k)}|\sqrt{p(m+k,m+k)}.\nonumber
\end{align}
With the help of Cauchy-Schwarz inequality, lower bound of right-hand-side of Eq.~(\ref{triangleineq}) is obtained.
\begin{align}\label{CSineq}
&\sum_{n}\left[A_{xm}^{2}B_{ym}^{2}\left|\sum_{i=0}^{2}\sqrt{p(i,i)}e^{i\zeta_{i}}\gamma_{ii}(n)\right|^{2}+\sum_{i=0}^{2}|A_{xm}B_{ym}-A_{xi}B_{yi}|^{2}p(i,i)|e^{i\zeta_{i}}\gamma_{ii}(n)|^{2}\right.\\
&\quad -\left.2A_{xm}B_{ym}\left|\sum_{i=0}^{2}\sqrt{p(i,i)}e^{i\zeta_{i}}\gamma_{ii}(n)\right|\sum_{k=1}^{2}Z_{xym}(k)|e^{i\zeta_{m+k}}\gamma_{(m+k)(m+k)}(n)|+2\prod_{k=1}^{2}D_{xym}(k)|e^{i\zeta_{m+k}}\gamma_{(m+k)(m+k)}|\right]\nonumber\\
&\geq \sum_{n}A_{xm}^{2}B_{ym}^{2}\left|\sum_{i=0}^{2}\sqrt{p(i,i)}e^{i\zeta_{i}}\gamma_{ii}(n)\right|^{2}+\sum_{i=0}^{2}|A_{xm}B_{ym}-A_{xi}B_{yi}|^{2}p(i,i)\sum_{n}|e^{i\zeta_{i}}\gamma_{ii}(n)|^{2}\nonumber\\
&\quad -2A_{xm}B_{ym} \sum_{k=1}^{2}D_{xym}(k)\sqrt{\sum_{n}|\sum_{i=0}^{2}\sqrt{p(i,i)}e^{i\zeta_{i}}\gamma_{ii}(n)|^{2}\sum_{n'}|e^{i\zeta_{m+k}}\gamma_{(m+k)(m+k)}(n')|^{2}}\nonumber\\
&=\left[A_{xm}B_{ym}\sqrt{\sum_{n}|\sum_{i=0}^{2}\sqrt{p(i,i)}e^{i\zeta_{i}}\gamma_{ii}(n)|^{2}}-\sum_{k=1}^{2}D_{xym}(k)\right]^{2}- 2\prod_{k=1}^{2}D_{xym}(k)\nonumber
\end{align}
because $\sum_{n=0}|\gamma_{xy}(n)|^{2}=1$. From Eq.~(\ref{nophaseineq}), Eq.~(\ref{triangleineq}) and Eq.~(\ref{CSineq}), we obtain the inequality
\begin{align}\label{beforegoal}
\left[\sqrt{p(\bar{x},\bar{y})}+\sum_{i\neq j}^{2}A_{xi}B_{yj}\sqrt{p(i,j)}\right]^{2} \geq& \left[A_{xm}B_{ym}\sqrt{\sum_{n}|\sum_{i=0}^{2}\sqrt{p(i,i)}e^{i\zeta_{i}}\gamma_{ii}(n)|^{2}}-\sum_{k=1}^{2}D_{xym}(k)\right]^{2} - 2\prod_{k=1}^{2}D_{xym}(k)
\end{align}
Rearranging Eq.~(\ref{beforegoal}) about $\sum_{n}|\sum_{i=0}^{2}\sqrt{p(i,i)}e^{i\zeta_{i}}\gamma_{ii}(n)|^{2}$, the inequality becomes
\begin{align}\label{finalineq}
\sum_{n}\left|\sum_{i=0}\sqrt{p(i,i)}e^{i\zeta_{i}}\gamma_{ii}(n)\right|^{2} \leq
\frac{1}{A_{xm}^{2}B_{ym}^{2}}\left\{\sqrt{\left[\sqrt{p(\bar{x},\bar{y})}+\sum_{i\neq j}A_{xi}B_{yi}\sqrt{p(i,j)}\right]^{2}+2\prod_{k=1}^{2}D_{xym}(k)}+\sum_{k=1}^{2}D_{xym}(k)\right\}^{2}
\end{align}
only when $A_{xm}B_{ym}\neq 0$. In order to simplify the description, the function $S$ is defined as
\begin{align}
S_{xy}(m)=\frac{2}{3A_{xm}^{2}B_{ym}^{2}\sum_{i,j=0}^{2}p(i,j)}\left\{\sqrt{\left[\sqrt{p(\bar{x},\bar{y})}+\sum_{i\neq j}A_{xi}B_{yj}\sqrt{p(i,j)}\right]^{2}+2\prod_{k=1}^{2} D_{xym}(k)}+\sum_{k=1}^{2}D_{xym}(k)\right\}^{2}.
\end{align}
The upper bound function $f$ and the factor $\varepsilon$ is defined with the function $S$ as explained in main text.
\end{widetext}
\end{appendix}

\end{document}